\RequirePackage{ifpdf}
\ifpdf
	\documentclass[showkeys, showpacs, pdftex, rmp]{revtex4}
\else
	\documentclass[showkeys, showpacs, dvips, rmp]{revtex4}
\fi

\ifpdf
  \usepackage[%
    pdftex,%
    pdftitle = {{Should there be more women in science and engineering?}},%
	pdfsubject = {{Everybody is so sure that asking whether there should be more women in science and engineering would be beating a dead horse that nobody really cared to check that the horse was actually dead.}},%
	pdfauthor = {{Mathieu Bouville}},%
	pdfkeywords = {{female students, higher education, university, ethics, policy, outreach programs, minority students}},%
	letterpaper,%
    hyperindex,%
    bookmarksopen,%
    bookmarksopenlevel=2,%
	breaklinks,%
]{hyperref}
\else
  \usepackage[
	dvips,
    pdftitle = {{Should there be more women in science and engineering?}},%
	pdfsubject = {{Everybody is so sure that asking whether there should be more women in science and engineering would be beating a dead horse that nobody really cared to check that the horse was actually dead.}},%
	pdfauthor = {{Mathieu Bouville}},%
	pdfkeywords = {{female students, higher education, university, ethics, policy, outreach programs, minority students}},%
	letterpaper,%
    hyperindex,%
    bookmarksopen,%
    bookmarksopenlevel=2,%
	breaklinks,%
]{hyperref}
\fi

\begin{document}

\title{\addcontentsline{toc}{chapter}{Should there be more women in science and engineering?}Should there be more women in science and engineering?}

\author{Mathieu Bouville}
	\email{m-bouville@imre.a-star.edu.sg}
	\affiliation{Institute of Materials Research and Engineering, Singapore 117602\addcontentsline{toc}{section}{Abstract}}

\begin{abstract}
Many people hold this truth to be self-evident, that there should be more female students in science and engineering. 
Typical arguments include possible benefits to women, possible benefits to the economy, and the unfairness of the current female under-representation. However, these justifications are never explicitly and thoroughly presented. Clearly stating and scrutinizing them, we show that they in fact have logical flaws. When made consistent, these arguments do not unconditionally justify enrolling more women in scientific disciplines. In particular, what women want must be taken into account.
Outreach programs towards \mbox{K--12} girls must therefore purport to allow them to choose a field freely, rather than try to draw as many of them to scientific disciplines as possible. This change of mindset must be accompanied by a close examination of the purpose and effects of these programs.
\end{abstract}
\keywords{female students, gender equity, higher education, university, ethics, policy, outreach programs, minority students}
\maketitle





\section{Introduction \texorpdfstring{---}{-} Raising a horse from the dead}
Far fewer women than men study science and engineering. It is often argued that women would benefit from graduating in a scientific discipline, due to higher salaries and the possibility to help others. Having more female engineers would also be beneficial to the economy because of the increasing need for engineers and of the positive impact of diversity on designs.
Another argument is that this under-representation of women is unfair. The question is then how to enroll more women in scientific disciplines.

In most articles on the subject, justifications for a greater female enrollment are relegated to the introduction (i.e.\ the only part of the article that needs not be in any way original). Some provide no justification at all or very vague ones, such as ``for a variety of practical and moral reasons'' \cite{Felder-95}. 
The closest authors get to presenting arguments is naming them: they mention the name of an argument \mbox{---r}ather than the argument it\mbox{self---,} say that it has been widely used (probably implying that it must therefore be valid), and move on. They for instance say ``a lot of people argue for diversity in terms of fairness \mbox{[ .\ .\ .\ ]} but that's not my argument''\ \cite{Wulf-diversity}, ``fairness is one answer, but certainly not the only one''\ \cite{Gosink-01}, or ``aside from the obvious issues of access, fairness and equity''\ \cite{Sullivan-03} without ever actually making these arguments explicit. Yet, such words do not imply the existence (let alone the validity) of arguments any more than dragons exist because the word `dragon' does.
In fact, everyone is so convinced that asking why there should be more female students in science and engineering would be beating a dead horse that nobody checked whether the horse was dead.

They would reply that their arguments need not be made explicit because they are obvious. Also `obvious' is the revolution of the sun around the earth. Philosopher Alfred North Whitehead said that ``it requires a very unusual mind to undertake the analysis of the obvious.'' Such a mind should not be very unusual amongst scientists and engineers. They should be able and willing to ``undertake the analysis of the obvious.'' In the present context, this requires a precise study of these justifications, even (or especially) when they seem obvious. The least this will take is to state arguments explicitly and clearly rather than glibly allude to them. Logic and accuracy are still invaluable and should still be used outside of science.

The objective of \citet{WISEST} at the University of Alberta is ``to investigate the reasons why few young women are choosing careers in the sciences and engineering [and] to take action to alter the situation.'' This means trying to solve a problem before knowing whether it exists at all. Yet, science and engineering have been successful in part because we first understand phenomena before using that knowledge for specific purposes. One would expect scientists to apply the high standards of science to the issue at hand: understanding should precede action. If an argument justifies a higher enrollment only under certain conditions, the actions one takes cannot contradict these conditions. Otherwise, the policy is literaly unjustified. As with a mathematical theorem, if the assumptions under which the argument was obtained do not hold, neither does the conclusion. Therefore, one cannot decorrelate justification and action.

Several kinds of arguments are often mentioned. While more arguments may strengthen one's position, it also means a greater probability that they be mutually exclusive. For instance, some argue that a greater female enrollment in science and engineering is good for the fields ---but not necessarily for the women themselves---, whereas to others it is a matter of freedom of choice. While each of these arguments might justify a greater female enrollment, they are obviously incompatible.

The purpose of this article is to analyze the arguments most commonly invoked to justify a greater female enrollment in science and engineering. In particular, one must make sure that they are self-consistent, that they do entail what they are supposed to prove, and that they are not incompatible. Philosopher and Nobel laureate Bertrand Russell pointed out that ``everything is vague to a degree you do not realize till you have tried to make it precise.'' We will try to make all arguments precise and oftentimes realize how vague (or ambiguous) they indeed were. Section~\ref{motivations} focuses on four arguments: higher salaries, the possibility to help others, the positive impact of diversity on designs, and the increasing need for engineers. Section~\ref{sec-fairness} is about fairness and under-representation. Section~\ref{sec-outreach} considers some consequences for policy of the arguments scrutinized in the previous two sections.

\section{\label{motivations}Mutual attraction}

\subsection{Motivating science and engineering}
\label{subsec-economy}
As technology plays an ever-increasing role in society and economy, the number of engineering graduates must increase. And since females and minorities are under-represented they offer great potential sources of engineers \cite{Rockland-02, Moskal-00, Chen-96, Zywno-99, Cuny-00, Baum-90, Brainard-98, Wulf-diversity, Sullivan-03, Grose-06, Lane-99}. Furthermore, more women in science and engineering would be good because of the greater variety of designs which more diverse teams could invent \cite{Wulf-diversity, Moskal-00, Gosink-01, Cuny-00, Lane-99}.

These arguments seem altruistic since women can \emph{contribute} to the economy: ``More compelling arguments have been raised that recognize the direct benefits that female participation is likely to have upon these fields.''\ \cite{Moskal-00}. But compelling for whom? These arguments may efficiently motivate science and engineering to attract women, but this is not sufficient. In essence, these arguments treat women as mere pawns to be transferred from one department to another based on some external reason (e.g.\ the economy). Can one force somebody to do something for the sake of others? According to Immanuel \citet{Kant-moral}, this is wrong because one should never treat people merely as means, but always as ends. To utilitarians, good means maximizing the happiness of the greatest number \cite{Bentham, Mill}; if women do not want to become engineers, having more of them in the field will decrease their happiness while increasing that of others only marginally. (One can notice that forbidding that students choose their field freely and getting the best students to fields which are deemed a priority by the State is typically a practice of dictatorships, not democracies.)

These economic arguments achieve the great fit of getting to agree (against them) ethical views which seldom do so: be it from a Kantian or a utilitarian viewpoint, these arguments must be rejected if women do not want to study science and engineering, i.e.\ they are not valid justifications on their own. 
That science and engineering should want to attract women does not imply that there \emph{has to} be more women in science and engineering: what women want is crucial.

Some disagree and argue that one needs more female engineers independently of what women want. Whether this is correct or not, one must at the very least be consistent and apply the same rule to all possible contributions to the economy. If child labor and child pornography can benefit the economy, should they be legalized? 
There are several ways out of this problem. One may ignore the issue altogether; this is ever popular yet not quite satisfying. One may stand by the premise of efficacy maximization and accept its logical consequences; this is self-consistent but will never get wide approval. One may claim that there exist particular cases; but a rule with too many exceptions is not a rule at all. One may provide a refined argument which does not endorse child labor and child pornography; but this path is at best narrow and it may not even exist. Finally, one may look for an argument of a completely different nature, which would not have the dreadful consequences of efficacy; this we will attempt in Sec.~\ref{sec-fairness}.

\subsection{\label{sec-motivation_women}Motivating women}
Another common argument is that ``the compensation in science-related fields is often higher than that of other fields. By not participating in science-related fields, women are barred from the economic rewards of these fields.''\ \cite{Moskal-00}. \Citet{Sullivan-03} make it clear early that engineering means money by paying high school girls to attend their summer engineering workshop.
However, if money were all that matters, why would there be engineering schools at all, given that engineering professors could earn more in industry? Apparently, women do not pick a field trying to maximize their income either: ``clearly, if monetary incentives were enough, current starting salaries would have already fixed the problem''\ \cite{Wulf-diversity}.

Women are more likely to stress interpersonal factors (e.g.\ helping others), whereas men tend to value money and status more \cite{Eccles-94, Morgan-01}. Would engineering allow women to help others? The president of the U.~S. National Academy of Engineering argues that engineering does have a positive impact on society and provides an opportunity to help others \cite{Wulf-diversity}. Let us assume for the sake of argument that he is right. What does this argument actually prove? It shows that women are wrong if they choose not to study engineering because they believe that it will not allow them to help others. That is, such an argument shows that women should want to major in engineering. It does not show that there has to be more women in engineering.

\subsection{Four shortcomings of the argument by mutual attraction}
We must first notice that taken in isolation demographics, better designs, higher salaries, and helping others all fail to justify anything. It is necessary that both women and the fields have an interest in an increased female enrollment, i.e.\ it is necessary to invoke these arguments together. Although having only one argument is not as satisfying as having four, it can be sufficient. But is this argument really satisfactory?

Do women want to marry engineering? They do. Does engineering want to marry women? It does. It is clear that they should marry. But if they decide not to, does one have a good reason to marry them against their will? 
Let us consider a simple analogy. We had two bottles of champagne and drank one, therefore there should be one left. Here `should' obviously has to do with a rational expectation based on the circumstances: it would not be immoral if it turned out that no champagne was left, it would be illogical. Something must be wrong with the premise: maybe we drank the second bottle or we had only one to begin with. Saying that since there should remain a bottle and none is to be found, we should go buy another one to solve the problem would be solving the wrong problem. 
Likewise, the fact that one would expect more women in science and engineering does not necessarily imply that one should increase female enrollment. In essence, this argument by mutual attraction tells us what to expect but does not provide us with any justification to make it happen if it does not spontaneously take place. In other words, `more women should be enrolled' does not imply `we should enroll more women.' As \citet{Appiah-05} points out, ``my life's shape is up to me \mbox{[ .\ .\ .\ ]}, even if I make a life that is less good than a life I could have made. All of us could, no doubt, have made better lives than we have: but that is no reason for others to attempt to force these better lives upon us.'' 
This impossibility to intervene makes this argument rather weak. 

Since the needs of the economy and the possibility to help others are not universal, neither is the win-win situation. For instance, it is doubtful that \emph{all} engineers have a positive impact on society: do weapons of war and buggy operating systems help others? Even if it is because of a small minority, the claim that engineering has a positive impact on society collapses: only certain jobs would be motivating, not any engineering job. Moreover, the need for engineers is not eternal (demand may decrease due to recession, outsourcing, etc.)\ and engineers from all fields are not wanted everywhere all the time.
The question `should there be more women in engineering?'\ is thus meaningless: there are as many questions as fields, times, places, etc. You can justify all female enrollment some of the time, and some female enrollment all the time, but you cannot justify all female enrollment all the time. What looked like a justification for a greater female enrollment in science and engineering is in fact career counseling.

Those arguing in favor of a greater female enrollment in scientific fields typically fail to notice that these extra female students must come from somewhere. But whence? If getting more female engineers means fewer female physicians, it is not necessarily an improvement (it may be bad for the population or for female representation in medicine). The question of women in science and engineering cannot be addressed in a vacuum: indirect effects  also matter. In mathematical terms: one needs to use differentials rather than partial derivatives. What is needed is a global rather than a local optimum; that more female engineers would be a good is irrelevant if more female physicians would be even better.

\subsection{Two notes on diversity}
\Citet{Anderson-02} want to ``increase diversity in engineering'' but never explain why diversity should be increased in the first place.
This is typical of a tendency to use `diversity' meta\-physically, taking it to be intrinsically valuable. Although `should there be more women?'\ and `should there be more diversity?'\ may seem equivalent in our context, the answer to the latter question is necessarily yes since diversity is presented as necessarily good.  But then, one should increase diversity in terms of men/women but also for blond/dark hair, short/tall, slim/fat, and even smart/stupid. Many will complain that these categories are completely irrelevant. But what is the criterion for relevance? For it to exist, there must be a reason why diversity is good (relevant meaning congruent with this reason). If there is no specific reason why diversity is good, there cannot be such a thing as irrelevant criteria. Clearly, the idea that diversity is good in and for itself is not tenable. Diversity is valuable only in its consequences and the actuality of these consequences must be explicitly shown.

One should also notice that women are not the only ones who can contribute to diversity. Foreigners (possibly hired due to the lack of local engineers) undoubtedly bring new ideas, which can lead to new designs. The low enrollment in engineering schools is therefore beneficial to society since it leads to better designs. This is another example of incompatible arguments: if diversity is crucial then low overall enrollment in scientific fields is a solution, not a problem.

\section{\label{sec-fairness}Fairness}
The argument by mutual attraction examined in the previous section turns out to be much weaker than its proponents believe. Another possible argument for an increased female enrollment is fairness. 
Typically, one just utters the word `fairness' and expects everybody to bow. There seems to be no need to clarify what one means by `fair' nor in what a higher female enrollment would be fair. It is common to use `fair' in a very loose way, as a synonym of `moral' or `good.' In this case, it has no precise meaning, just a vague positive value: `we should do what is fair' then means `we should do what is good', i.e.\ `we should do what we should do.' This tautology is obviously useless. For fairness to be of any use, it must have a precise meaning.

\subsection{The superiority of science and engineering}
Another way to construe fairness is to say that science and engineering being superiorly good, everybody should have access to them. This is true of freedom, justice, education, etc. Yet, this requires that these fields be indeed superior. Based on some criteria they may be, but then one needs to show that these criteria are intrinsically superior to any other, with the risk of an infinite regression. Moreover, most fields are probably convinced of their own superiority (like most religions are convinced that they are the one true religion) and there is no reason to assume that science and engineering are right and everybody else is wrong. 
So far, no demonstration of the superiority of scientific disciplines has been provided. The burden of proof is on those who hold that all fields are not equally acceptable career choices (just like one assumes that people are equal and demands that anyone claiming otherwise provide a proof that all humans are not equal.)

\subsection{\label{sec-Hume}Under-representation: statistical viewpoint}
It is taken for granted that women are under-represented in science and engineering and that their enrollment should therefore be increased \cite{Baum-90, Chen-96, Brainard-98, Cohoon-02, Zywno-99, Anderson-02, Roberts-02, Sullivan-03, Chubin-05}.
Yet, as Russell pointed out, ``in all affairs it's a healthy thing now and then to hang a question mark on the things you have long taken for granted.'' 

As a matter of fact, ``while women make up over 50\% of the college-age population in the U.S., they represent a small minority among engineering students''\ \cite{Chen-96}. Yet, does it really imply that it is ``critically important that the causes for the low enrollment and high attrition rates of female engineering students be identified and eliminated''\ \cite{Chen-96}? This claim relies on the simplest and most common of arguments based on under-representation: having no argument at all \cite{Baum-90, Chen-96, Brainard-98, Cohoon-02, Zywno-99, Morgan-01, Anderson-02, Roberts-02, Sullivan-03, Chubin-05}. (1) Women are under-represented. (2) Their enrollment ought to be increased. (3) They are not under-represented any more. QED.

Yet, under-representation is of a statistical nature (lower proportion of women than in the overall population), it is neither good nor bad \emph{per se} \cite{Bouville-Phys_World-07}. Despite what \citet{Chubin-05} believe, ``improving'' or ``promoting'' the number of women does not have any meaning whatsoever. The claim that ``the numbers speak for themselves, demonstrating a significant problem in recruiting and retaining women.''\ \cite{Baum-90} is ventriloquism. If under-representation were wrong in itself then the under-representation of women in prison would imply that there should be more female inmates. It should now be obvious that one cannot simply say without any explanation that because there are few women in science and engineering there ought to be more (saying that there are \emph{too few} women would already assume that there ought to be more).  As \citet{Hume-treatise} pointed out, one must be very careful when trying to derive an `ought' from an `is.' If one means to equate under-representation and unfairness, a link between the two must be provided explicitly.

\subsection{\label{underrepresentation-unfairness}Under-representation: ethical viewpoint}
Under-representation taken in this crude statistical sense is obvious but irrelevant. Some, such as \citet{Cuny-00}, seem to vaguely see that the problem cannot be purely statistical as ``underrepresentation translates into a \emph{loss of opportunity} for individuals'' (emphasis added), yet they do not grasp the implications since they still aim for ``an effort by all departments to increase the total number of women.'' For under-representation to be a valid argument, it indeed must be (re)defined to be ethical, rather than statistical, in nature. The way to do this is to change the reference from the average population to the situation in a perfectly fair world, i.e.\ a world devoid of prejudice, discrimination, and other biases. Women are then said to be under-represented if the level of female enrollment in the actual world is lower than what it would be in a perfectly fair world. This new definition is less straightforward than comparing enrollment to the proportion of women in the overall population but it can link under-representation to unfairness, which the latter cannot do. In particular, it does not justify an increase of the number of female inmates, which is obviously an improvement.
 
One speaks of equal opportunity when people who have equivalent abilities and who perform an equivalent amount of work reach equivalent results. In a perfectly fair world, men and women would obviously have equal opportunities. What about the real world? The low enrollment of females in science and engineering indicates that they do not reach similar results. But do men and women have similar abilities? Moreover, there can be unfairness only if women who want to study science are barred from the field, not if low enrollment springs from low interest. We must therefore examine both the abilities and motivations of women compared to men.

\subsection{Sexual differences in abilities and occupational interests}
Due to the greater variability of males compared to females, men are over-represented at the top \cite{Hedges-95} and at the bottom  of intellectual ability, e.g.\ four times as many boys as girls are dyslexic \cite{Kleinfeld-98}.
Moreover, men and women have different strengths: women typically have better verbal skills and men mathematical and spatial abilities. This difference is (at least in part) hormonal and has been observed in non-human species \cite{Browne-05}.
Men and women do not differ only in terms of abilities, they also make different choices. On Holland's vocational interest test, women score higher on the ``artistic'' and ``social'' dimensions and men on the ``investigative'' and ``realistic'' (relevant to science and engineering) dimensions \cite{Kaufman-98}.
Part of the difference is biological: many sex-differentiated behaviors appear at an early age when children are unable to identify sexes \cite{Serbin-01, Connellan-00} and sexual differences of taste have been found in other species \cite{Alexander-02}.
\label{sec-career_choice}
Part of the difference is due to socialization: parents, teachers, and the students themselves typically have lower expectations in math for girls compared to boys \cite{Jacobs-85, Hyde_etal-90}. 

While social biases would not exist in a perfectly fair world, biological differences would. Hence, all we can conclude is that the fair level of enrollment is above the current figure (which is too low due to social biases) but below 50\% (innate differences prevent this).
While this conclusion seems anticlimactic, any claim beyond this (e.g.\ that the fair level of enrollment is close to 50\%) would be mere opinion, as it would \emph{assume} that one of the two contributions dominates.

\subsection{Fairness and freedom of choice}
The ethical redefinition of under-representation is an improvement but does not suffice. If all women studied science then obviously there would be no under-representation. However, this would be far from ideal since it would require forcing many women to study science against their will. Indeed, under-representation ---even redefined ethically--- is intrinsically asymmetric: it can see when there is too little but is blind to cases of too much. A symmetric criterion would say that women should be neither barred from nor forced to the field. One can notice that this is not based on averages but on individuals. In effect, this means giving up the superfluous concept of under-representation altogether, which was a bad way of referring to an actual problem. We will simply say that fairness exists when all women can graduate in a field congruent with their abilities and desires. One must also notice that enrollment numbers too are irrelevant, since they take women to be undistinguishable. 

Should there be more women in science and engineering? No, there should only be more women in science and engineering who want to be in these fields. The argument of Sec.~\ref{motivations} based on efficacy and the fairness argument we just considered lead to surprisingly similar conclusions. We can therefore be confident that our conclusion that every woman should be allowed to graduate in a field congruent with her abilities and desires is robust. We did find a good reason to increase female enrollment in science and engineering, but it applies only under precise conditions.

\section{\label{sec-outreach}Implications for outreach programs}

As we have seen, many people argue that there should be more female students in science and engineering. The next step they take is to try and increase female enrollment in these fields. A popular solution takes the form of outreach programs towards K--12 girls (generally in high schools but not only), giving them a more hands-on experience of science and engineering \cite{Anderson-02, Rockland-02, Moskal-00, Muller-97, Cohoon-02, Zywno-99, Baum-90, Sullivan-03}.

While the primary aim of this article is to analyze arguments in favor of a greater female enrollment in science and engineering, it seems natural to also look at their consequences for outreach programs. Yet, exactly how the collapse of the usual justifications affects outreach will be for its practitioners to determine: 
\emph{they} are in the best position to evaluate and alter the programs in which they are involved. This section can but provide some possible leads.

\subsection{\label{more_better}The more the better}
Outreach programs are considered ``recruitment'' and their success is measured in terms of number of participants who eventually enroll in science and engineering \cite{Baum-90, Brainard-98, Cohoon-02, Moskal-00, Cuny-00, Gosink-01, Rockland-02, Anderson-02, Roberts-02, Sullivan-03, Grose-06, Lane-99, Chubin-05, Zywno-99}. Clearly, this equates more with better: the purpose is to get as many women as possible to scientific fields, independently of what these women want. Bean counters may parade as grand liberators, but even when the rhetoric mentions a free career choice ---e.g.\ ``women pursuing their \emph{interests} in science'' \cite[][emphasis added]{Muller-97}---, success is measured in terms of enrollment numbers. Free the beans! 

One must keep in mind that the reason why outreach programs were created in the first place was that the number of females in science and engineering was artificially low, due to the impossibility for girls to freely choose a career in  these fields. Outreach programs tend towards the opposite extreme: an impossibility for girls to freely choose \emph{against} a career in science or engineering. It would certainly be paradoxical to trample their right to a free choice in order to enforce it. 
Sharing one's love for science and engineering and trying to cancel out a negative bias are very different from preaching the science gospel and actively converting girls. Manipulating a girl towards science is not any more acceptable than manipulating her away from it.

Some insist that many outreach programs are voluntary. Yet, one cannot say that outreach is justified because girls cannot choose freely their career and that outreach cannot be an issue because the girls freely choose to attend. Either girls can choose freely and outreach is pointless or girls are not free to choose and outreach can indeed create a new servitude. In either case, outreach programs shortchange girls.

\subsection{Who makes the choice?}
Some will also counter that none of the authors cited actually argued that women should be forced into science and engineering. Yet, none argued that women should \emph{not} be forced into these disciplines either. Nor do they at any time explain how they try to prevent this from happening. The question of freedom of choice thus seems morally neutral to them (while the claim may not be explicit, it is embodied in their actions). Since they do not see freedom of choice as relevant, they cannot take it into account.

Studies show that ``people form enduring interests in activities in which they view themselves to be self-efficacious'' yet ``people may rely more on perceived than tested abilities in formulating their interests.''\ \cite{Lent-94}. By showing girls that they can be successful in science or engineering, one does not expunge past influences to increase their freedom of choice. Rather one generates a new influence, this time in favor of the scientific disciplines. 
It is not possible to erase social biases, one can only overwrite them: outreach is bound to tamper with the girls' interests and manipulate girls towards scientific fields.

Take three girls. Tell the first one how rewarding engineering is and show her that she can be a successful engineer. Do the same with the other two girls and medical and law schools. What will happen? The first girl will graduate in engineering, the second in medicine, and the third in law. Does the outcome correlate with what girls want? Not at all. It correlates with the viewpoint that they have heard. \emph{We} are the ones making a decision by choosing what information we give them.

\subsection{Reinventing outreach}
While the idea that more is better is clearly mistaken, finding a practical replacement is not straight\-forward: obviously, assessing the match between a (possibly unexpressed) preference and a career is more difficult than assessing enrollment. Just as obviously, one should never set a goal just based on how easily success can be evaluated.

Were we to tell girls all they need to know about the many alternatives (science, engineering, law, medicine, business, journalism, teaching, art, staying at home, etc.), they may be able to make a personal informed choice. Those who believe that engineering has mostly an image problem \cite[e.g.][]{Wulf-diversity} would expect this to `solve the problem' since, even if competing choices are also presented in a fair way, engineering would benefit the most. If they are right then the issue of low female enrollment can be resolved without trampling the rights of the girls to choose freely. One can notice that this is not science and engineering outreach anymore, but rather fair and neutral career counseling.

One must not overestimate the impact of the rejection of the typical arguments. Indeed, an invalid demonstration does not make the theorem one is trying to prove false. Strictly speaking, the fact that none of the arguments supposed to justify a greater female enrollment actually does does not prove that outreach programs are an error. What it does show is that there currently is no valid justification for the outreach programs in their current form. One may either revise them to make them consistent with what justifies them (i.e.\ make them justified) or one may propose a new argument proving their validity. In any case, one cannot continue programs devoid of rational justification.

\section{Conclusion}
Many in science and engineering hold this truth to be self-evident, that there should be more women in the field. We considered several commonly proposed justifications: higher salaries, the possibility to help others, the increasing need for engineers, and the impact of diversity on new designs. When made explicit and seriously scrutinized, they in fact show that there should be a mutual attraction between women and scientific fields. This attraction is not universal so that there should be more women in science and engineering only inasmuch as women actually want to graduate in these fields.

Many claim that women are under-represented in science and engineering and that, out of fairness, their enrollment should be increased. Yet, under-representation is statistical in nature, it is neither right nor wrong. We therefore redefined it to make it an ethical concept. Although an improvement, this still had a shortcoming: its asymmetry. We finally settled on a criterion which did not mention under-representation and simply states that all women should be allowed to graduate in a field congruent with their abilities and desires. This conclusion is similar to that obtained from mutual attraction.

As currently conceived, outreach means trying to sell our product (science and engineering) to as many potential female customers as possible. It does not aim at empowering women to choose freely any more than any other sales pitch does. Since drawing always more women to science and engineering violates their right to choose a career freely, a change of mindset ---a `paradigm shift'--- from increasing enrollment to increasing freedom is necessary. 
Since present outreach is incompatible with its justifications, programs have to be modified to be really justified (or better arguments must be proposed to justify the programs in their current form). 
Of course, this is easier said than done: trying to allow women to choose a field freely would be a moral dream but also a practical nightmare. Nevertheless, one cannot go on undisturbed on a path which appears to be the wrong one.

\end{document}